\ifx\onecol\undefined
\documentclass[journal]{IEEEtran}
\else 
\documentclass[onecolumn,draftcls,12pt]{IEEEtran}
\fi

\usepackage[utf8]{inputenc}
\usepackage{physics}
\usepackage{amsmath}
\usepackage{amssymb}
\usepackage{subfigure}
\usepackage{graphicx}
\usepackage{graphics}
\usepackage{color}
\usepackage{psfrag}
\usepackage{cite}
\usepackage{balance}
\usepackage{algorithm}
\usepackage{accents}
\usepackage{amsthm}
\usepackage{bm}
\usepackage{url}
\usepackage{algorithmic}
\usepackage[english]{babel}
\usepackage{multirow}
\usepackage{enumerate}
\usepackage{cases}
\usepackage{stfloats}
\usepackage{dsfont}
\usepackage{soul}
\usepackage{amsfonts}
\usepackage{fancyhdr}
\usepackage{hhline}
\usepackage{array}
\usepackage{booktabs}
\usepackage{framed}
\usepackage{float}
\usepackage{soul}
\usepackage{url}
\usepackage{hyperref}

\newcommand{\mb}[1]{{  \mathbf  #1}}  

\begin{document}
\newtheorem{theorem}{Theorem}
\newtheorem{acknowledgement}[theorem]{Acknowledgement}
\newtheorem{axiom}[theorem]{Axiom}
\newtheorem{case}[theorem]{Case}
\newtheorem{claim}[theorem]{Claim}
\newtheorem{conclusion}[theorem]{Conclusion}
\newtheorem{condition}[theorem]{Condition}
\newtheorem{conjecture}[theorem]{Conjecture}
\newtheorem{criterion}[theorem]{Criterion}
\newtheorem{definition}{Definition}
\newtheorem{exercise}[theorem]{Exercise}
\newtheorem{lemma}{Lemma}
\newtheorem{corollary}{Corollary}
\newtheorem{notation}[theorem]{Notation}
\newtheorem{problem}[theorem]{Problem}
\newtheorem{proposition}{Proposition}
\newtheorem{solution}[theorem]{Solution}
\newtheorem{summary}[theorem]{Summary}
\newtheorem{assumption}{Assumption}
\newtheorem{example}{\bf Example}
\newtheorem{remark}{\bf Remark}

\newtheorem{thm}{Corollary}[section]
\renewcommand{\thethm}{\arabic{section}.\arabic{thm}}

\def\qed{$\Box$}
\def\QED{\mbox{\phantom{m}}\nolinebreak\hfill$\,\Box$}
\def\proof{\noindent{\emph{Proof:} }}
\def\poof{\noindent{\emph{Sketch of Proof:} }}
\def
\endproof{\hspace*{\fill}~\qed
\par
\endtrivlist\unskip}
\def\endproof{\hspace*{\fill}~\qed\par\endtrivlist\vskip3pt}

\def\E{\mathsf{E}}
\def\eps{\varepsilon}
\def\phi{\varphi}
\def\Lsp{{\boldsymbol L}}
\def\Bsp{{\boldsymbol B}}
\def\lsp{{\boldsymbol\ell}}
\def\Ltsp{{\Lsp^2}}
\def\Lpsp{{\Lsp^p}}
\def\Linsp{{\Lsp^{\infty}}}
\def\LtR{{\Lsp^2(\Rst)}}
\def\ltZ{{\lsp^2(\Zst)}}
\def\ltsp{{\lsp^2}}
\def\ltZt{{\lsp^2(\Zst^{2})}}
\def\ninN{{n{\in}\Nst}}
\def\oh{{\frac{1}{2}}}
\def\grass{{\cal G}}
\def\ord{{\cal O}}
\def\dist{{d_G}}
\def\conj#1{{\overline#1}}
\def\ntoinf{{n \rightarrow \infty}}
\def\toinf{{\rightarrow \infty}}
\def\tozero{{\rightarrow 0}}
\def\trace{{\operatorname{trace}}}
\def\ord{{\cal O}}
\def\UU{{\cal U}}
\def\rank{{\operatorname{rank}}}
\def\acos{{\operatorname{acos}}}

\def\SINR{\mathsf{SINR}}
\def\SNR{\mathsf{SNR}}
\def\SIR{\mathsf{SIR}}
\def\tSIR{\widetilde{\mathsf{SIR}}}
\def\Ei{\mathsf{Ei}}
\def\l{\left}
\def\r{\right}
\def\lb{\left\{}
\def\rb{\right\}}

\setcounter{page}{1}

\newcommand{\eref}[1]{(\ref{#1})}
\newcommand{\fig}[1]{Fig.\ \ref{#1}}

\def\bydef{:=}
\def\ba{{\mathbf{a}}}
\def\bb{{\mathbf{b}}}
\def\bc{{\mathbf{c}}}
\def\bd{{\mathbf{d}}}
\def\bee{{\mathbf{e}}}
\def\bff{{\mathbf{f}}}
\def\bg{{\mathbf{g}}}
\def\bh{{\mathbf{h}}}
\def\bi{{\mathbf{i}}}
\def\bj{{\mathbf{j}}}
\def\bk{{\mathbf{k}}}
\def\bl{{\mathbf{l}}}
\def\bm{{\mathbf{m}}}
\def\bn{{\mathbf{n}}}
\def\bo{{\mathbf{o}}}
\def\bp{{\mathbf{p}}}
\def\bq{{\mathbf{q}}}
\def\br{{\mathbf{r}}}
\def\bs{{\mathbf{s}}}
\def\bt{{\mathbf{t}}}
\def\bu{{\mathbf{u}}}
\def\bv{{\mathbf{v}}}
\def\bw{{\mathbf{w}}}
\def\bx{{\mathbf{x}}}
\def\by{{\mathbf{y}}}
\def\bz{{\mathbf{z}}}
\def\b0{{\mathbf{0}}}

\def\bA{{\mathbf{A}}}
\def\bB{{\mathbf{B}}}
\def\bC{{\mathbf{C}}}
\def\bD{{\mathbf{D}}}
\def\bE{{\mathbf{E}}}
\def\bF{{\mathbf{F}}}
\def\bG{{\mathbf{G}}}
\def\bH{{\mathbf{H}}}
\def\bI{{\mathbf{I}}}
\def\bJ{{\mathbf{J}}}
\def\bK{{\mathbf{K}}}
\def\bL{{\mathbf{L}}}
\def\bM{{\mathbf{M}}}
\def\bN{{\mathbf{N}}}
\def\bO{{\mathbf{O}}}
\def\bP{{\mathbf{P}}}
\def\bQ{{\mathbf{Q}}}
\def\bR{{\mathbf{R}}}
\def\bS{{\mathbf{S}}}
\def\bT{{\mathbf{T}}}
\def\bU{{\mathbf{U}}}
\def\bV{{\mathbf{V}}}
\def\bW{{\mathbf{W}}}
\def\bX{{\mathbf{X}}}
\def\bY{{\mathbf{Y}}}
\def\bZ{{\mathbf{Z}}}

\def\bxi{{\boldsymbol{\xi}}}

\def\sT{{\mathsf{T}}}
\def\sH{{\mathsf{H}}}
\def\cmp{{\text{cmp}}}
\def\cmm{{\text{cmm}}}
\def\WPT{{\text{WPT}}}
\def\lo{{\text{lo}}}
\def\gl{{\text{gl}}}

\def\tT{{\widetilde{T}}}
\def\tF{{\widetilde{F}}}
\def\tP{{\widetilde{P}}}
\def\tG{{\widetilde{G}}}
\def\tbh{{\widetilde{\mathbf{h}}}}
\def\tbg{{\widetilde{\mathbf{g}}}}

\def\mA{{\mathbb{A}}}
\def\mB{{\mathbb{B}}}
\def\mC{{\mathbb{C}}}
\def\mD{{\mathbb{D}}}
\def\mE{{\mathbb{E}}}
\def\mF{{\mathbb{F}}}
\def\mG{{\mathbb{G}}}
\def\mH{{\mathbb{H}}}
\def\mI{{\mathbb{I}}}
\def\mJ{{\mathbb{J}}}
\def\mK{{\mathbb{K}}}
\def\mL{{\mathbb{L}}}
\def\mM{{\mathbb{M}}}
\def\mN{{\mathbb{N}}}
\def\mO{{\mathbb{O}}}
\def\mP{{\mathbb{P}}}
\def\mQ{{\mathbb{Q}}}
\def\mR{{\mathbb{R}}}
\def\mS{{\mathbb{S}}}
\def\mT{{\mathbb{T}}}
\def\mU{{\mathbb{U}}}
\def\mV{{\mathbb{V}}}
\def\mW{{\mathbb{W}}}
\def\mX{{\mathbb{X}}}
\def\mY{{\mathbb{Y}}}
\def\mZ{{\mathbb{Z}}}

\def\cA{\mathcal{A}}
\def\cB{\mathcal{B}}
\def\cC{\mathcal{C}}
\def\cD{\mathcal{D}}
\def\cE{\mathcal{E}}
\def\cF{\mathcal{F}}
\def\cG{\mathcal{G}}
\def\cH{\mathcal{H}}
\def\cI{\mathcal{I}}
\def\cJ{\mathcal{J}}
\def\cK{\mathcal{K}}
\def\cL{\mathcal{L}}
\def\cM{\mathcal{M}}
\def\cN{\mathcal{N}}
\def\cO{\mathcal{O}}
\def\cP{\mathcal{P}}
\def\cQ{\mathcal{Q}}
\def\cR{\mathcal{R}}
\def\cS{\mathcal{S}}
\def\cT{\mathcal{T}}
\def\cU{\mathcal{U}}
\def\cV{\mathcal{V}}
\def\cW{\mathcal{W}}
\def\cX{\mathcal{X}}
\def\cY{\mathcal{Y}}
\def\cZ{\mathcal{Z}}
\def\cd{\mathcal{d}}
\def\Mt{M_{t}}
\def\Mr{M_{r}}
\def\O{\Omega_{M_{t}}}
\newcommand{\figref}[1]{{Fig.}~\ref{#1}}
\newcommand{\tabref}[1]{{Table}~\ref{#1}}

\newcommand{\fb}{\tx{fb}}
\newcommand{\nf}{\tx{nf}}
\newcommand{\BC}{\tx{(bc)}}
\newcommand{\MAC}{\tx{(mac)}}
\newcommand{\Pout}{p_{\mathsf{out}}}
\newcommand{\nnn}{\nn\\}
\newcommand{\FB}{\tx{FB}}
\newcommand{\TX}{\tx{TX}}
\newcommand{\RX}{\tx{RX}}
\renewcommand{\mod}{\tx{mod}}
\newcommand{\m}[1]{\mathbf{#1}}
\newcommand{\td}[1]{\tilde{#1}}
\newcommand{\sbf}[1]{\scriptsize{\textbf{#1}}}
\newcommand{\stxt}[1]{\scriptsize{\textrm{#1}}}
\newcommand{\suml}[2]{\sum\limits_{#1}^{#2}}
\newcommand{\sumlk}{\sum\limits_{k=0}^{K-1}}
\newcommand{\eqhsp}{\hspace{10 pt}}
\newcommand{\tx}[1]{\texttt{#1}}
\newcommand{\Hz}{\ \tx{Hz}}
\newcommand{\sinc}{\tx{sinc}}
\newcommand{\diag}{\mathrm{diag}}
\newcommand{\MAI}{\tx{MAI}}
\newcommand{\ISI}{\tx{ISI}}
\newcommand{\IBI}{\tx{IBI}}
\newcommand{\CN}{\tx{CN}}
\newcommand{\CP}{\tx{CP}}
\newcommand{\ZP}{\tx{ZP}}
\newcommand{\ZF}{\tx{ZF}}
\newcommand{\SP}{\tx{SP}}
\newcommand{\MMSE}{\tx{MMSE}}
\newcommand{\MINF}{\tx{MINF}}
\newcommand{\RC}{\tx{MP}}
\newcommand{\MBER}{\tx{MBER}}
\newcommand{\MSNR}{\tx{MSNR}}
\newcommand{\MCAP}{\tx{MCAP}}
\newcommand{\vol}{\tx{vol}}
\newcommand{\ah}{\hat{g}}
\newcommand{\tg}{\tilde{g}}
\newcommand{\teta}{\tilde{\eta}}
\newcommand{\heta}{\hat{\eta}}
\newcommand{\uh}{\m{\hat{s}}}
\newcommand{\eh}{\m{\hat{\eta}}}
\newcommand{\hv}{\m{h}}
\newcommand{\hh}{\m{\hat{h}}}
\newcommand{\Po}{P_{\mathrm{out}}}
\newcommand{\Poh}{\hat{P}_{\mathrm{out}}}
\newcommand{\Ph}{\hat{\gamma}}
\newcommand{\mat}[1]{\begin{matrix}#1\end{matrix}}
\newcommand{\ud}{^{\dagger}}
\newcommand{\C}{\mathcal{C}}
\newcommand{\nn}{\nonumber}
\newcommand{\nInf}{U\rightarrow \infty}

\title{Rydberg Atomic Receiver: Next Frontier of Wireless Communications}

\author{{Mingyao Cui, Qunsong Zeng, and Kaibin Huang,~\IEEEmembership{Fellow, IEEE}}
\thanks{The authors are with the Department of Electrical and Electronic Engineering, The University of Hong Kong, Hong Kong. Corresponding author: K. Huang (Email: huangkb@eee.hku.hk).}}



\maketitle


\begin{abstract}
Rydberg Atomic REceiver (RARE) is driving a paradigm shift in electromagnetic wave measurement by harnessing the electron transition phenomenon of Rydberg atoms.
Operating at the quantum scale, such receivers have the potential to breakthrough the performance limit of classic receivers, sparking a revolution in physical-layer wireless communications.  
The objective of this paper is to offer insights into RARE-empowered communication systems. We first provide a comprehensive introduction to the fundamental principles of RAREs. Then, a thorough comparison between RAREs and  classic receivers is conducted in terms of the antenna size, sensitivity, and bandwidth. Subsequently, we overview the recent progresses in RARE-aided wireless communications, covering the frequency-division
multiplexing, multiple-input-multiple-output, wireless sensing, and quantum many-body techniques. 
{\color{black} Moreover, the unique application of RARE in multi-band communication and sensing is unveiled, followed by numerical experiments to show its performance superiority over classic receivers by $\sim2.4\:{\rm bps/Hz}$ in spectrum efficiency and  $\sim7.2\:{\rm dB}$ in sensing accuracy.} 
Finally, we conclude this paper by providing promising research directions. 
\end{abstract}

\begin{IEEEkeywords}
Rydberg atomic receivers, wireless communications, quantum sensing.
\end{IEEEkeywords}

\section{Introduction}

The six-generation (6G) communication networks will leverage all available resources to support the explosive growth of global mobile services.
To achieve seamless coverage,  6G will expand beyond terrestrial networks in 1G–5G to space–air–ground integrated networks. The full spectrum, including sub-6G, millimeter-wave (mmWave), and terahertz (THz) bands, will be explored for attaining high-rate and multi-modal connections. 
These ambitious visions require the next-generation receiver to exhibit exceptional sensitivity to electromagnetic (EM) waves across a wide spectrum. 

In current communication systems, receivers detect wireless signals by using an antenna to transform incident EM waves into electric currents~\cite{meyer_assessment_2020}. 
This mechanism, however, presents insurmountable limitations to physical-layer (PHY) wireless communications. Firstly, the front-end circuits of a receiver, like amplifiers and mixers, introduce thermal noise into data currents, restricting the receiver's sensitivity limit. In addition, an ideal dipole antenna typically has a half-wavelength size, which constrains its working bandwidth to a narrow range around the central frequency. To detect signals from sub-6G to THz, the receiver has to equip an array of bulky antennas and font-end circuits, each dedicated to a specific band. Consequently, these challenges constrain the performance upper-bound of wireless communications~\cite{zhang_rydberg_2024}.

Recently, at the intersection of quantum sensing and wireless communication, \emph{Rydberg Atomic REceiver (RARE)} has emerged as a radical solution to overcome these limitations~\cite{fan_atom_2015}. 
Rydberg atoms are highly-excited atoms with outer electrons far from the nucleus. Due to the large electric dipole moments, Rydberg atoms can strongly interact with incident EM waves, triggering electron transitions between energy levels. As a result, RAREs can accurately detect wireless signals by monitoring changes on the atoms' quantum states resulting from electron transitions~\cite{zhang_rydberg_2024}. 
Representing a paradigm shift in receiver architecture to embrace quantum sensing, RAREs are gaining increasing attentions due to their quantum advantages. 
For instance, a RARE is immune to thermal noise as it involves no front-end circuits. Hence, its sensitivity limit can potentially exceed those of classic counterparts~\cite{QuanSense_Zhang2023}.
Moreover, a single RARE is able to interact with a large number of frequencies ranging from Megahertz (MHz) to THz band because of the abundant energy levels in Rydberg atoms~\cite{RydMultiband_Meyer2023}. 
These quantum advantages make RAREs a promising technology to revolutionize the next-generation wireless network. 
The objective of this paper is to provide a comprehensive guide for beginners to the RARE-empowered communication systems. To this end, we first introduce the fundamental characteristics  of Rydberg atoms and the key components of a RARE. Next, the key indicators of RAREs are compared with classic receivers in  terms of antenna size, sensitivity, and bandwidth. 
Subsequently, we review the state-of-the-art RARE techniques for wireless communications, covering frequency-division multiplexing (FDM), multiple-input-multiple-output (MIMO), quantum wireless sensing, and quantum many-body techniques. {\color{black} We then reveal a unique application of RAREs in multi-band communication and sensing systems---a single RARE can simultaneously detect communication signals at low frequencies to ensure coverage and sensing signals at high frequencies to enhance granularity. Numerical experiments confirm that RAREs outperform conventional counterparts, achieving a spectrum efficiency gain of $\sim2.4\:{\rm bps/Hz}$ and a sensing accuracy improvement of $\sim7.2\:{\rm dB}$. 
Last, we highlight promising future research directions on RARE from the PHY wireless communication perspective, including signal modeling, holographic atomic-MIMO,  satellite communications, and related areas.}

\section{Fundamentals of Rydberg Atomic Receivers}

\begin{figure*}
   \centering
   \includegraphics[width=7in]{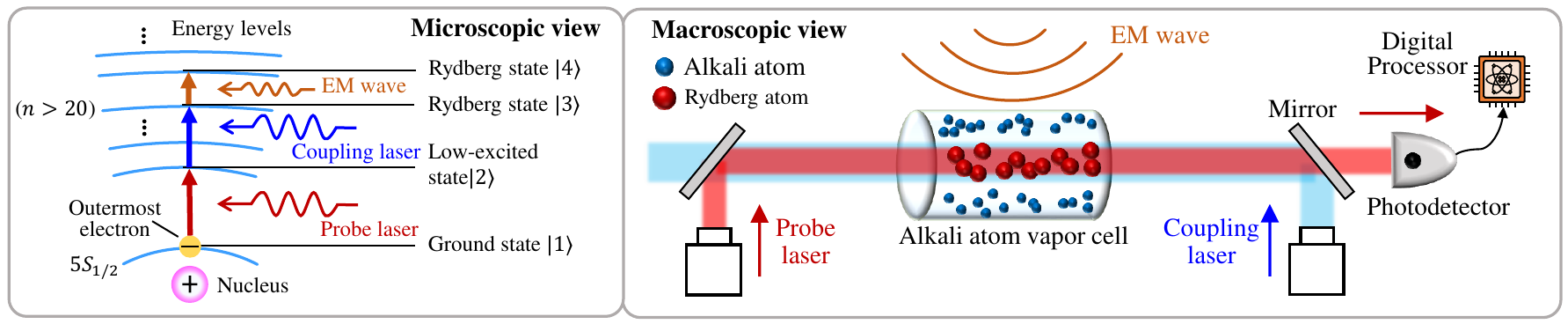}
   \vspace*{-1em}
   \caption{Framework of RARE-enabled wireless communications. The ``Microscopic view" adopts the structure of Rb atom as an example. 
   }
   \label{img:atomic receiver}
   \vspace*{-1em}
\end{figure*}

We commence by introducing the fundamentals of RAREs, ranging from the preliminary of Rydberg atom to the key modules of a RARE.

\subsection{Preliminary of Rydberg Atom}
\subsubsection{Definition of Rydberg atom}
An atom consists of a positively charged nucleus orbited by negatively charged electrons. Electrons in these orbits have unique quantum states and quantized energy levels. Each orbit is labeled by three quantum numbers $n\ell_j$, (e.g., $n\ell_j = 5S_{1/2}$), including a principal quantum number $(n = 1, 2, 3, \cdots)$, an azimuthal quantum number $(\ell = S, P, D, \cdots)$, and a total angular momentum quantum number $(j = \frac{1}{2}, \frac{3}{2}, \frac{5}{2},\cdots)$. 
While the latter two numbers, $\ell$ and $j$, determine the angular momentum and orbital shape, the principal quantum number, $n$, affects the orbital radius and energy. For a  higher value of $n$, the electron is farther from the nucleus and has higher energy (see Fig.~\ref{img:atomic receiver}). \emph{Rydberg atom is defined as a highly-excited atom with one or more electrons that have a large principal quantum number, i.e., $n>20$}. 
\subsubsection{Electron transition for preparing Rydberg atom} Electrons can transit between energy levels by absorbing photons from (or emitting photons to) EM waves~\cite{QuanSense_Zhang2023}.
This electron-transition process occurs when the photon's frequency resonates with the \emph{transition frequency between energy levels}, which is defined as the ratio between the energy-level spacing and Plank constant. 
\emph{The process facilitates the creation of Rydberg atoms via the excitation of electrons to high energy levels.} This is typically accomplished by counter-propagating two lasers, namely a probe laser and a coupling laser, through a vapor cell filled with Alkali atoms (e.g., Rb and Cs), as shown in Fig.~\ref{img:atomic receiver}~\cite{fan_atom_2015}.
The probe laser excites the outermost electrons from the ground state to a lowly-excited energy level, (e.g., $5S_{1/2}\rightarrow5P_{3/2}$). The coupling beam further triggers the electron transition to a Rydberg state, (e.g. $5P_{3/2} \rightarrow 47S_{5/2}$), thereby creating Rydberg atoms. Customarily, the three involved energy levels, (e.g., $5S_{1/2}$, $5P_{3/2}$, $47S_{5/2}$), are denoted as $\ket{1}$, $\ket{2}$, and $\ket{3}$~\cite{zhang_rydberg_2024}.
\subsubsection{Properties of Rydberg atom}
Given the high principal quantum number $n$, Rydberg atoms exhibit useful properties for sensing EM waves. First, the radius of an Rydberg atom is large due to its dependence on $n^2$~\cite{zhang_rydberg_2024}. For instance, the atomic radius grows from $150a_0$ to $10^4a_0$ as $n$ increases from $10$ to $100$, with $a_0 = 5.29\times 10^{-11}{\rm m}$ being the Bohr radius. The substantial increase in radius endows the Rydberg atom with a large electric dipole moment, thus making it very sensitive to external EM wave. Second, 
the transition frequency between adjacent Rydberg states is abundant as the energy level spacing scales as $n^{-3}$~\cite{fan_atom_2015}. For low energy levels ($n < 20$), the transition frequency can reach the visible light frequency ($>100$ THz). While for Rydberg states ($n > 20$), the transition frequencies fall within the range of MHz to THz~\cite{QuanSense_Zhang2023}, exactly covering all the typical frequency bands used in PHY communications.


\subsection{Key Modules of RAREs}
The extreme sensitivity of Rydberg atoms to EM waves ranging from MHz to THz motivates the development of RAREs for EM-wave measurement. A RARE typically comprises three key modules: electron-transition based antennas, \emph{Electromagnetically-Induced-Transparency (EIT)} based quantum measurement, and receive signal processing.

\subsubsection{Electron-transition based antennas}
The electron transitions induced by probe/coupling lasers and incident EM waves, provides the fundamental mechanism of RARE. Rydberg atoms, created by the probe/coupling lasers, are stored in a vapor cell to serve as a receive antenna (see Fig.~\ref{img:atomic receiver}).
The incident EM wave, $\Re{\mb{E}(t)e^{j2\pi ft}}$, interacts with these atoms and excite their electrons to transit from the initial Rydberg state $\ket{3}$ to a new Rydberg state $\ket{4}$, where the transition frequency $f_{\ket{3} \rightarrow  \ket{4}}$ is properly selected to resonate with the radio frequency $f$. During this process, the information carried by the EM wave is transduced into the quantum states.

To be more precise, the quantum state of this four-level system ($\ket{1}\sim\ket{4}$) is represented by a $4\times 4$ \emph{density matrix} $\boldsymbol{\rho}$. 
The electron transition drives $\boldsymbol{\rho}$ to evolve according to the \emph{Lindblad master equation}~\cite{RydNP_Jing2020}, i.e., Schrödinger equation in open quantum systems. 
When the system becomes steady, \emph{the EM wave will affect the final quantum state $\boldsymbol{\rho}(\Omega(t))$ via a physical parameter $\Omega(t)$ called Rabi frequency}~\cite{jia_properties_2024}. Rabi frequency characterizes the interaction strength between the electric dipole moment $\boldsymbol{\mu}$ and EM wave, which is expressed as  
$\Omega(t) = \frac{1}{\hbar}\sqrt{\left|{\boldsymbol{\mu}^T \mb{E}(t)}\right|^2 + \hbar^2\delta^2}$~\cite{AtomicMIMO_Cui2024}, 
where $\hbar$ denotes the reduced Plank constant and $\delta$ the detuning of radio frequency $f$ from the transition frequency $f_{\ket{3}\rightarrow\ket{4}}$. Two crucial conclusions can be drawn from $\Omega(t)$.  
First, Rydberg atoms act as both an antenna and a down-converter, as the electron transition, $\ket{3} \leftrightarrow \ket{4}$, inherently down-converts the EM wave, $\Re{\mb{E}(t)e^{j2\pi ft}}$, to the baseband signal, $\mb{E}(t)$. 
Second, the signal, $\mb{E}(t)$, is amplified by the electric dipole moment, $\boldsymbol{\mu}$. Therefore, a higher Rydberg state will contribute to an increased sensitivity to EM waves~\cite{fan_atom_2015}.


\subsubsection{EIT based quantum measurement}
\begin{figure}
   \centering
   \includegraphics[width=3.5in]{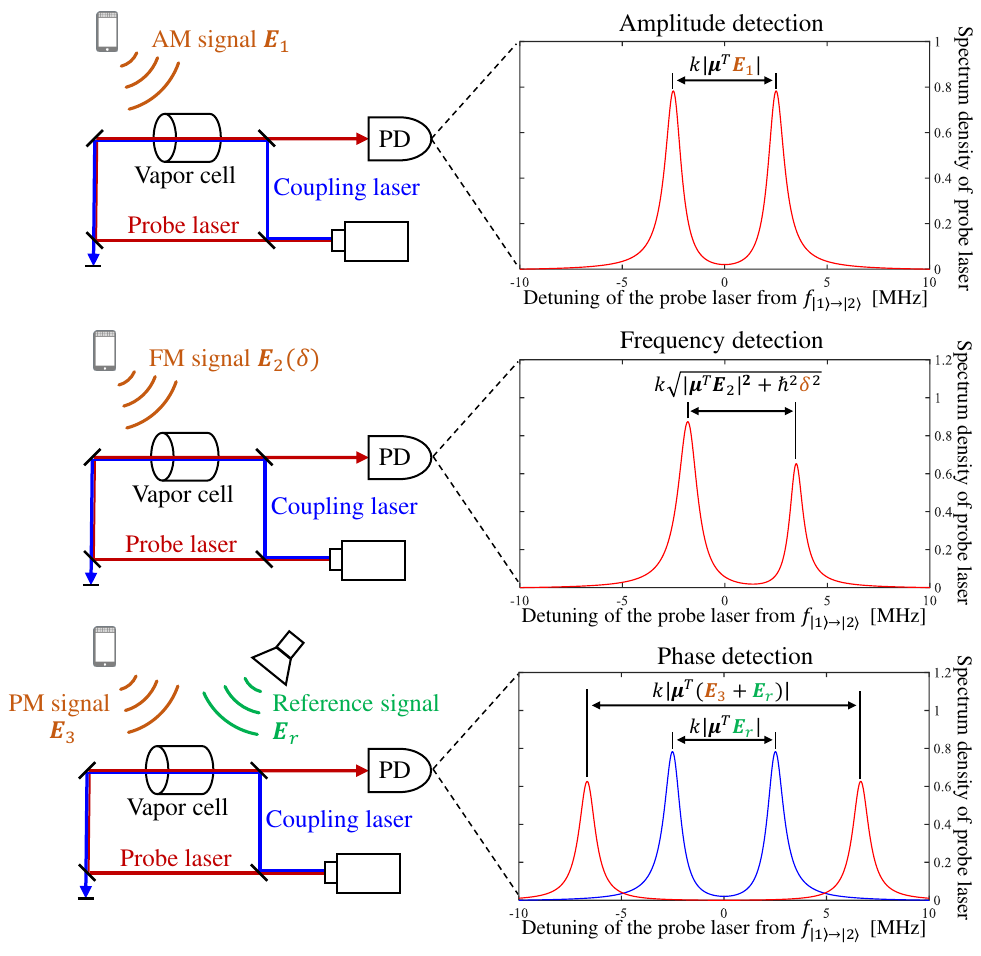}
   \vspace*{-1em}
   \caption{EIT for measuring AM/FM/PM signals. For AM and PM detection, the detuning $\delta$ is zero, while for FM detection,  $\delta$ is nonzero. 
   }
   \label{img:EIT}
   \vspace*{-1em}
\end{figure}
To retrieve the wireless signal, it is essential to accurately readout the value of Rabi frequency. The standard measurement scheme is the EIT spectroscopy,  developed on the following observations~\cite{sedlacek_microwave_2012}.
\begin{itemize}
    \item The $(1,2)$-th element of the final quantum state, $\rho_{12}$, determines the rate at which Rydberg atoms absorb photons from the probe laser. The absorption loss of the probe-laser scales exponentially with $\rho_{12}$~\cite{RydNP_Jing2020}. Thereby, it is feasible to deduce the value of $\rho_{12}$ by measuring the output power of the probe laser using a photodetector (PD), 
    as depicted in Fig.~\ref{img:atomic receiver}.
    \item The quantum state, $\rho_{12}$, has a one-to-one mapping with the Rabi frequency. Specifically, $\rho_{12}$ displays two near-zero points at two specific probe-laser frequencies situated around the transition frequency $f_{\ket{1} \rightarrow \ket{2}}$. The interval of these two frequencies is shown to be linearly proportional to the Rabi frequency $\Omega(t)$~\cite{fan_atom_2015}.
\end{itemize} 
These observations suggest that by sweeping the probe-laser frequency around $f_{\ket{1} \rightarrow \ket{2}}$, two peaks can appear in the spectrum of the output probe laser, as depicted in Fig.~\ref{img:EIT}.  This allows the Rabi frequency to be measured from the peak interval, which is the principle of EIT-based quantum measurement. 

\subsubsection{Receive signal processing} After obtaining the value of Rabi frequency, the information modulated on $\mb{E}(t)$ can be recovered via customized signal processing implemented on traditional digital processors. As in Fig.~\ref{img:EIT}, both amplitude-modulated (AM) and frequency-modulated (FM) symbols can be illustrated directly using the maximum likelihood method due to the dependence of Rabi frequency on $\left|{\boldsymbol{\mu}^T \mb{E}(t)}\right|$ and $\delta$. 
For the detection of phase-modulated (PM) symbols,  
a heterodyne sensing scheme was developed in~\cite{RydNP_Jing2020}, which originates from the holographic-phase sensing technique in optics imaging. It introduces an auxiliary source to produce \emph{a known reference signal}, $\mb{E}_r(t)$, which interferes with the unknown data signal, $\mb{E}(t)$, over the air.  
Their radio frequencies can be either identical or a few kHz apart. 
Consequently, the measured Rabi frequency changes from $|\boldsymbol{\mu}^T\mb{E}(t)|$ to $|\boldsymbol{\mu}^T\mb{E}(t) + \boldsymbol{\mu}^T\mb{E}_r(t)|$ (see the bottom-subfigure in Fig.~\ref{img:EIT}). Accordingly, the information modulated on the global phase of $\mb{E}(t)$ can be extracted from the relative phase between $\mb{E}(t)$ and $\mb{E}_r(t)$ using spectrum analysis methods or phase retrieval algorithms~\cite{AtomicMIMO_Cui2024}.

So far, we have discussed the mechanism of detecting AM/FM/PM wireless signals using RAREs. Recent studies have also reported the detection of multi-frequency signals~\cite{liu_deep_2022}, multi-user signals~\cite{AtomicMIMO_Cui2024}, and target sensing~\cite{QuanSense_Zhang2023, QWC_Cui2025}, all based on the principles we have covered  and will be elaborated in Section~\ref{sec:4}.

\section{Rydberg Atomic Receiver vs. Classic Receiver} 
The RARE technique represents a paradigm shift in wireless signal detection from using electronic circuits to embracing quantum sensing. This gives rise to a range of unique features for RARE-enabled PHY wireless communications, which are introduced in this section.

\subsection{Antenna Size}
According to the Chu limit, a lossless and impedance-matched antenna in a conventional receiver requires a physical dimension proportional to the wavelength, e.g. $\frac{\lambda}{2}$, so as to efficiently convert EM waves into signals on a transmission line. At low frequencies, the size of an antenna element can reach hundreds of centimeters, for example exceeding $150\:{\rm cm}$ when $f<100\:{\rm MHz}$, leading to bulky receiver architectures. 
Conversely, a RARE does not absorb energy from the field, but rather leverages the atom-light interaction to perform nondestructive sensing. This mechanism enables the development of electrically small Rydberg antennas that are significantly smaller than the wavelength. Typically, the geometry size of a Rydberg antenna---comprising an ensemble of Rydberg atoms---is on the order of $1\:{\rm cm}\times 1\:{\rm mm}$, with the vapor cell length around $1\:{\rm cm}$ and the laser diameter approximately $1\:{\rm mm}$. This wavelength-independent size offers substantial advantages, especially for low-frequency applications.

\subsection{Sensitivity Limit}

The sensitivity of a receiver, $E_{\min}$, defined as the minimum detectable field strength per square root of averaging time 
(unit: $\rm V/m/\sqrt{Hz}$), 
is fundamentally limited by the internal noise generated by the receiver itself. 
For an ideal $\frac{\lambda}{2}$ dipole antenna of a classic receiver, its internal noise stems from the \emph{thermal noise} generated by front-end circuits, such as amplifiers and mixers.
The corresponding sensitivity limit is expressed as $E_{\min} =  \sqrt{\frac{P_N Z_0}{A_e}}$, where $P_N = -174{\rm dBm}/{\rm Hz}$ denotes the room-temperature thermal-noise power, $Z_0$ the free-space impedance, and $A_e$ the effective area. On the contrary, since a RARE contains no front-end circuits, its thermal noise is significantly weaker than a classic receiver~\cite{QuanSense_Zhang2023}. Its major internal noise is the \emph{quantum shot noise (QSN)} resulting from the randomness when observing quantum states~\cite{meyer_assessment_2020}. According to the quantum sensing theory, the QSN power is inversely proportional to the electric dipole moment, $\boldsymbol{\mu}$, the number of participating atoms, $N_a$, and the coherence time, $T_r$ of the EIT process. The theoretical sensitivity limit of a RARE, also called Standard Quantum Limit (SQL), is calculated as 
 $E_{\min} =\frac{\hbar}{|\boldsymbol{\mu}|\sqrt{N_aT_r}}$~\cite{meyer_assessment_2020}.

 \begin{figure}
   \centering
   \includegraphics[width=3.3in]{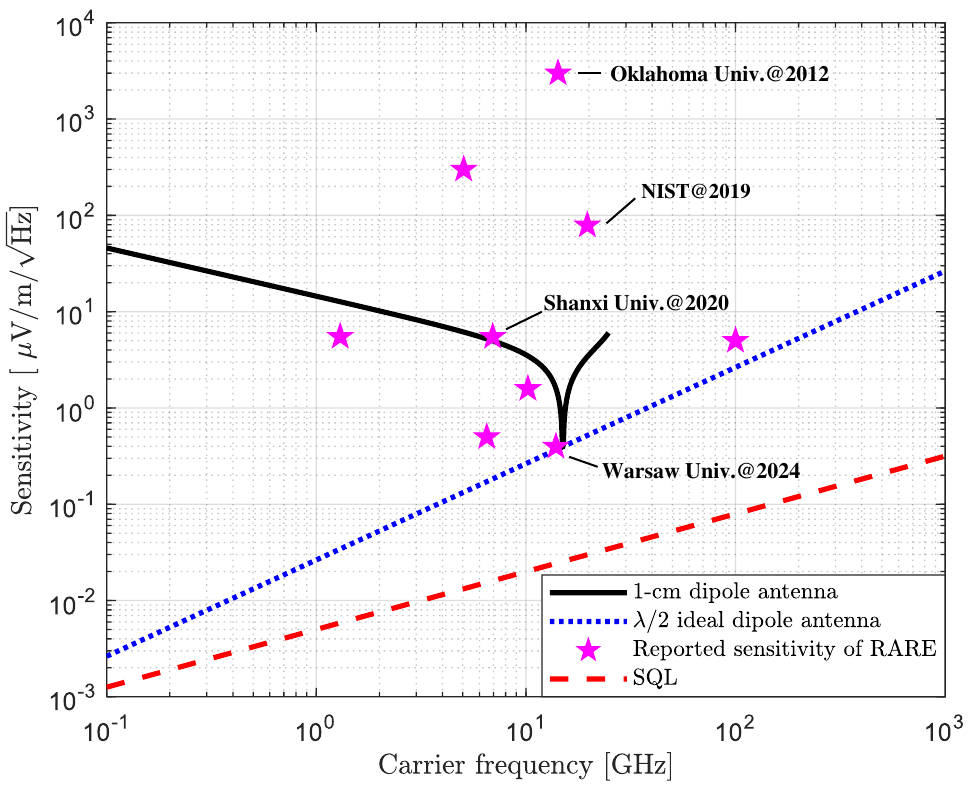}
   \vspace*{-1em}
   \caption{Sensitivity comparison between  RAREs and classic receivers.}
   \label{img:sensitivity}
   \vspace*{-1em}
\end{figure}
 Fig.~\ref{img:sensitivity} compares the sensitivity of RAREs with classic receivers.
Representative values $T_r = 225{\rm \mu s}$ and $N_a = 5\times10^5$ as in~\cite{fan_atom_2015} are utilized. The  dipole moment, $\boldsymbol{\mu}$, is selected from the electron transition $nD_{5/2}\leftrightarrow(n+1)P_{3/2}$ that resonates with each frequency. For a fair comparison, we also plot the sensitivity limit of a 1-cm dipole antenna calculated in \cite{meyer_assessment_2020} as it has a similar size with a vapor cell. Fig.~\ref{img:sensitivity} illustrates that the SQL can surpass the sensitivity limit of $\frac{\lambda}{2}$ dipole antennas by orders of magnitude. This is attributed to the large electric dipole moment of Rydberg atoms. Moreover, since the early exploitation of RARE in~\cite{sedlacek_microwave_2012} (Oklahoma Univ.@2012), the RARE's sensitivity has been improved by $10^4$ times. Recently developed RAREs have outperformed similarly sized dipole antennas and achieved sensitivities comparable to ideal $\frac{\lambda}{2}$ counterparts. The state-of-the-art design in~\cite{OpticalConverter_Sebastian2024} (Warsaw Univ.@2024) pushes the sensitivity down to $0.4 {\rm \mu V/m/\sqrt{Hz}}$ at 13.9 GHz by using six-wave-mixing techniques to remove background noises. 
It is anticipated that in the near future, RAREs will surpass the ideal $\frac{\lambda}{2}$ dipole antenna and approach SQL, with multi-wave-mixing emerging as a promising approach for performance improvement.

\subsection{Bandwidth}
 In existing RARE studies, the term "bandwidth" is often used to represent two distinct metrics. Firstly, it is utilized to define the range of frequencies that can trigger electron transitions of Rydberg atoms. This concept is commonly referred to as the ``\emph{detectable frequency range}" of a RARE in the literature. Secondly, it is also associated with the ``\emph{instantaneous bandwidth}", which denotes the maximum data capacity that the receiver device can measure at any given moment. This term aligns with the standard definition of "bandwidth" in communication engineering. 
Typically, a RARE is characterized by a broad detectable frequency range but a narrow instantaneous bandwidth.


\subsubsection{Detectable frequency range} The detectable frequency range of a RARE can be significantly broader than that of a classic receiver. This is attributed to two key factors. 
Firstly, Rydberg atoms possess abundant energy levels. This property allows electrons to interact with various frequency bands, ranging from MHz to THz bands, via the transition between different energy levels~\cite{zhang_rydberg_2024}.  For instance, transitions $56D_{5/2}\leftrightarrow57P_{3/2}$ and $56D_{5/2}\leftrightarrow52F_{3/2}$ can resonate with the radio frequencies 12.01~GHz and 116.05~GHz, respectively.
Furthermore, some quantum techniques, like the Zeeman effect and off-resonate ac Stark shift, can broaden the linewidth of electron transition. In this context, the detectable frequency range of a RARE can be further expanded to achieve continuous-frequency detection exceeding 1 GHz. 



\subsubsection{Instantaneous bandwidth} 
Restricted by the long response time for an EIT process to become steady, the instantaneous bandwidth of a RARE is typically narrower than 10 MHz~\cite{jia_properties_2024}. This bandwidth is comparable to that of 4G LTE systems (1$\sim$20 MHz).
Recent efforts are endeavored to increase the RARE's bandwidth. For example, researchers have proposed a spatiotemporal multiplexing technique of probe lasers to decrease the EIT response time (see [139] of \cite{zhang_rydberg_2024}). This technique expands the RARE's bandwidth to 100 MHz, approaching that of sub-6G systems specified by 5G standards (i.e., 20$\sim$100 MHz). To date, continuously improving RARE's instantaneous bandwidth remains a hot topic.


\section{State-of-the-art RARE Techniques}\label{sec:4}
There have been studies to develop advanced RARE techniques to enhance the communication and sensing performance.
This section provides an overview of the most recent progresses of these state-of-the-art designs. 

\subsection{Frequency-division Multiplexed RAREs}
The FDM technology plays a crucial role in 5G wideband systems for mitigating multi-path interference. The advancement of FDM-RARE is regarded as the key enabler of RARE-empowered wideband communications. The state-of-the-art FDM-RAREs can be separated into two categories: \emph{standard} and \emph{multi-band} FDM-RAREs.

\subsubsection{Standard FDM-RARE}
As illustrated in Fig.~\ref{img:OFDM_CBC}, a standard FDM-RARE aims to detect standard FDM signals,  
where all sub-carrier frequencies are compacted within a bandwidth of several MHz. In this context, a uniform electron transition 
 $\ket{3} \leftrightarrow \ket{4}$ is sufficient to down-convert all sub-carrier signals into the Rabi frequency of quantum state. The major challenge in facilitating standard FDM-RAREs is that the nonlinearity of Rabi frequency, $\Omega(t)$, w.r.t the EM wave destroys the orthogonality among different sub-carriers, resulting in significant intercarrier crosstalk.
To address this issue, a deep learning model was trained in~\cite{liu_deep_2022} to directly infer FDM symbols from the received probe-laser power. The experiment therein considered four subcarriers with a frequency spacing of 2~kHz in line-of-sight environment. It was shown that the learning approach can demodulate binary phase-shift keying symbols with over 99\% accuracy. 

\subsubsection{Multi-band FDM-RARE} 
\begin{figure}
   \centering
   \includegraphics[width=3.5in]{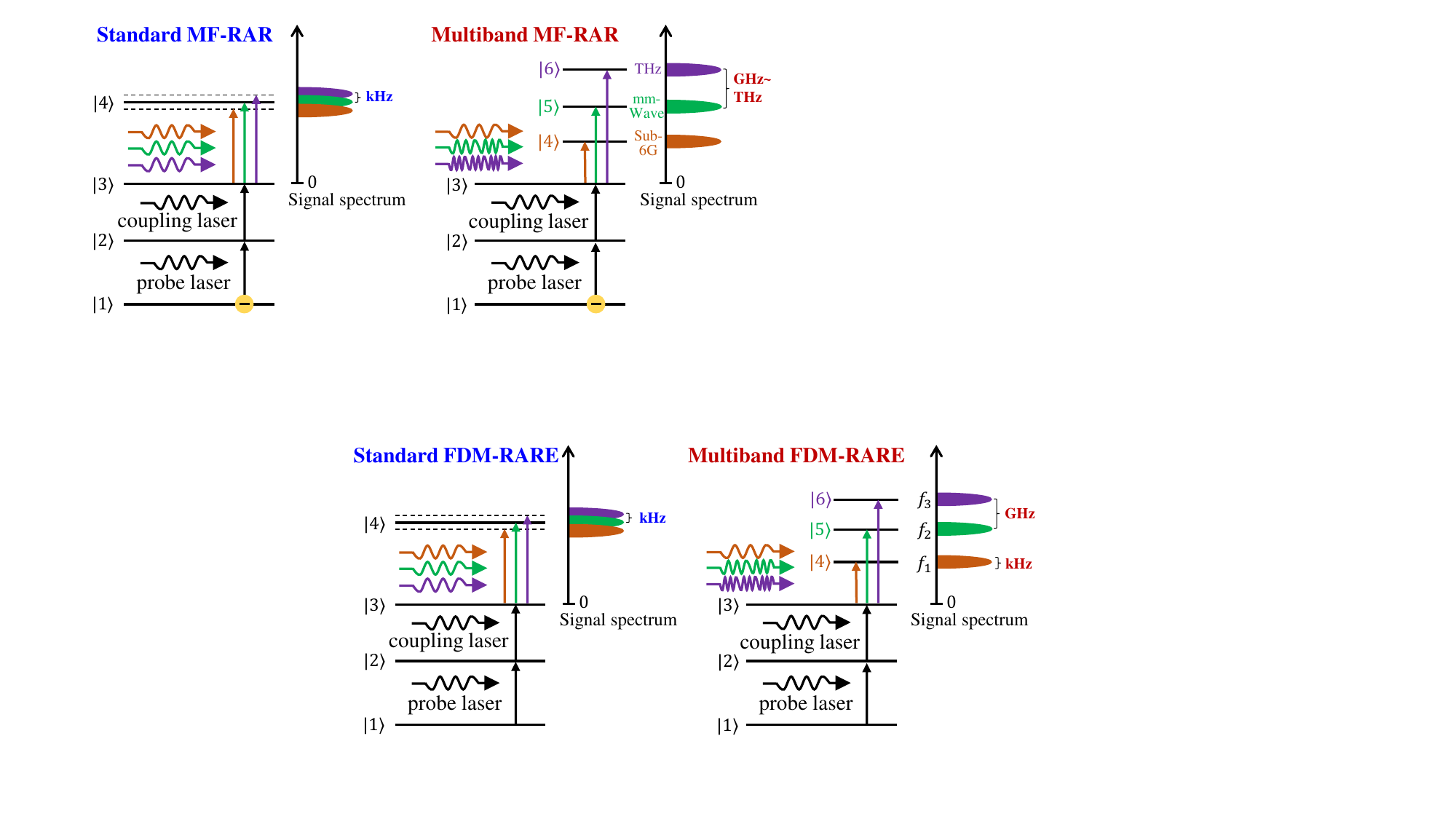}
   \vspace*{-1em}
   \caption{Standard and multi-band FDM-RAREs. 
   }
   \label{img:OFDM_CBC}
   \vspace*{-1em}
\end{figure}

The multi-band FDM-RARE leverages the broad detectable frequency range of a RAREs to serve as an integrated full-frequency communication platform.
On this platform, signal frequencies can be distributed across distinct frequency bands, with band spacing ranging from GHz to THz, as presented in Fig.~\ref{img:OFDM_CBC}.
To enable signal detection, the RARE needs to couple each band to a unique energy-level pair, denoted as $\ket{3} \leftrightarrow \ket{n}, n\in\{4,5,\cdots, N\}$.
The electron transition between each energy-level pair down-converts the coupled band by its respective transition frequency $f_{\ket{3}\rightarrow\ket{n}}$, thereby simultaneously converting the entire multi-band RF signal to the baseband.
Utilizing this principle, researchers have demonstrated the co-detection of 5 separate frequency bands (1.72, 12.11, 27.42, 65.11, and 115.75 GHz) with a 3-dB instantaneous bandwidth of 6 MHz, verifying RARE's full-frequency sensing capability~\cite{RydMultiband_Meyer2023}.  

\subsection{Atomic-MIMO Communications}
{\color{black} As a core technology in 5G and beyond,  MIMO communications can greatly enhance the spectral efficiency through the diversity and multiplexing gains. The recent study in \cite{FSO-MIMO_Elasyed2024} has demonstrated a 20~Gbps 1500~m transmission in free-space-optical communications using MIMO diversity combining codes.} 
The advancement of MIMO thus motivates the research on atomic-MIMO communications. Its key idea is to employ an array of RAREs, in place of traditional antenna arrays, for signal detection. In this context, both the high sensitivity of Rydberg atoms and the high spectral efficiency of MIMO can be leveraged. 


In the literature, a simpler atomic \emph{single-input-multiple-output} (SIMO) system was developed in \cite{otto_data_2021} for the first time to explore the diversity gain, wherein four RAREs were used to jointly recover AM symbols from a single-antenna user. 
Experiments validated that the receive SNR scales linearly with the number of RAREs, as in traditional systems.
{\color{black} To further reap the multiplexing gain, \cite{AtomicMIMO_Cui2024} has investigated the multi-user signal detection capability of atomic-MIMO.  
It was proved that atomic-MIMO exhibits a nonlinear \emph{phase-retrieval} transmission model. The nonlinear characteristic negates the efficacy of traditional zero-forcing and minimum-mean-square-error linear detectors, posing a challenge to atomic-MIMO signal detection. To this end, an Expectation-Maximization Gerchberg-Saxton algorithm was proposed to iteratively retrieve multi-user signals with a low complexity, demonstrating the feasibility of atomic-MIMO.}


\subsection{RARE-empowered Quantum Wireless Sensing}
Given that a RARE is capable of detecting extremely weak signals, it can be deployed in wireless sensing systems to refine the sensing accuracy by orders of magnitude. Specifically, one existing prototype senses a target's vibration from both WiFi (2.4 GHz) and mmWave (28 GHz) signals \cite{QuanSense_Zhang2023}. Therein, the heterodyne sensing scheme, as illustrated in Fig.~\ref{img:EIT}, was used to recover the phase of an EM wave and then retrieve the subtle-vibration of target. Experiment results show that the RARE can increase the
sensing accuracy by more than 10 times as compared with conventional receivers implemented on the USRP, LimdSDR, and Intel 5300 WiFi card. 

In addition, RARE can also be utilized to sense angle-of-arrivals (AoA) and time-of-arrivals (ToA) for the purpose of localization. A self-heterodyne sensing technique was introduced in \cite{QWC_Cui2025} to advance the current heterodyne sensing for high-resolution localization. This technique leverages the self-interference produced by the transmitted sensing signal as the reference signal, ensuring their waveforms to be identical. By doing so, it effectively overcomes the narrow instantaneous bandwidth limitation of RARE, enhancing the sensing bandwidth up to 100 MHz. Experiments show that a self-heterodyne RARE reduces the localization error by orders of magnitude compared with classic receivers. 


\subsection{Quantum Many-body RAREs for External-noise Mitigation}
A RARE is characterized by low internal noise, as previously discussed. However, it is still affected by external noises resulting from the inter-user and inter-cell interferences. This problem introduces an obstacle to the sensitivity advantage of RAREs. In recent years, several studies have been carried out to mitigate the external noise. 

One intriguing approach leverages the quantum many-body effect of Rydberg atoms to counterintuitively convert the power of external noise into that of wireless signal, achieving a \emph{noise-enhanced RARE} \cite{NonlinearRAR_Wu2024}. In the approach, a collection of Rydberg atoms are interacted with each other. This many-body interaction transforms the original \emph{linear} Schrödinger equation into a \emph{nonlinear} one, which is shown to exhibit a bistability  behavior. The incoming wireless signal could stimulate the system to switch between the two stable states.  Then, by introducing a certain amount of external noise into the receiver, the switching magnitude can be amplified such that the output SNR is drastically improved. Experiment results demonstrate that the noise-enhanced RARE can surpass the sensitivity of traditional RAREs by 6.6 dB in the presence of external noise. This work opens a new direction for advancing RARE: the detrimental  impact of external noise can be turned to a beneficial effect through the quantum many-body interactions.


{\color{black}
\section{RARE for Multi-band Sensing and Communication (RARE-MSAC)}

\begin{figure}
   \centering
   \includegraphics[width=3.5in]{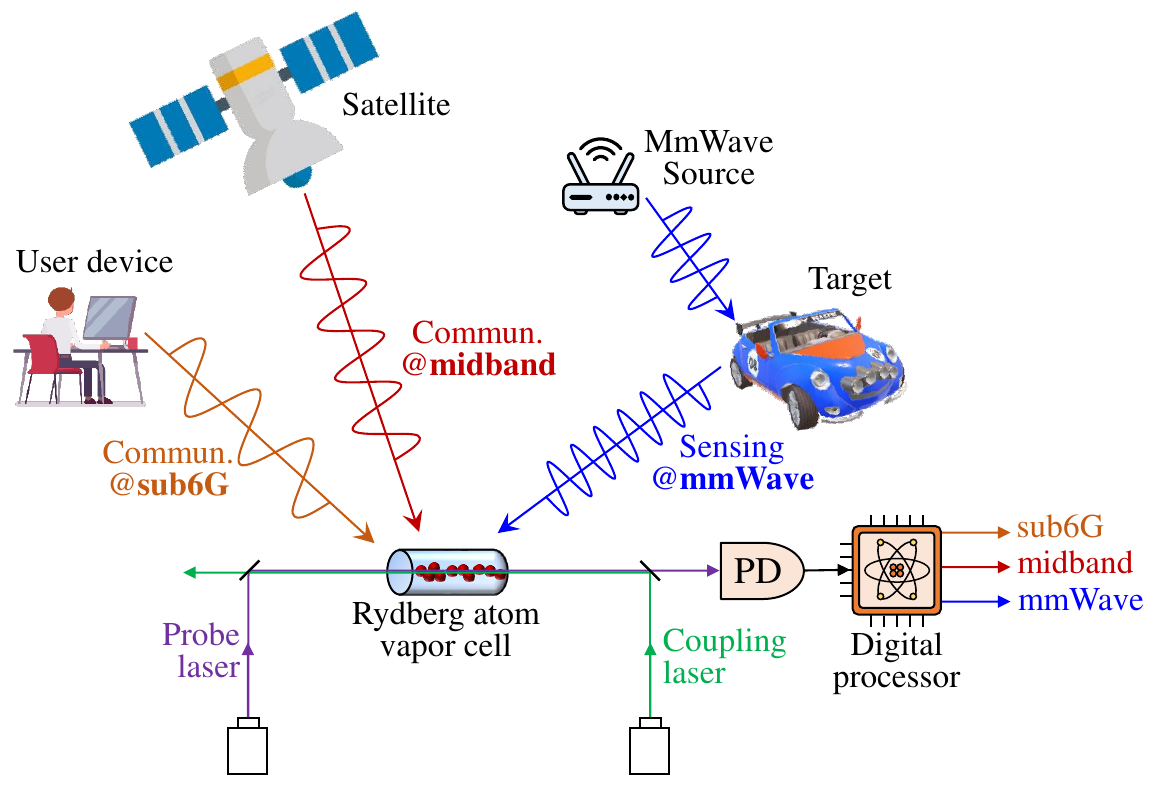}
   \vspace*{-1em}
   \caption{A single RARE is used to simultaneously detect communication and sensing signals at different frequency bands.  
   }
   \label{img:MSAC}
   \vspace*{-1em}
\end{figure}

We leverage the advancement of RARE to introduce a novel application in wireless systems------the RARE-MSAC. 
The key idea is to exploit the multi-level electron transitions of Rydberg atoms for detecting multi-band signals with diverse sensing and communication (S\&C) requirements.  As illustrated in Fig.~\ref{img:MSAC}, a \emph{single} RARE is employed. The atoms are prepared in an appropriate initial Rydberg state, $\ket{3}$, where the transitions from $\ket{3}$ to other Rydberg states, $\ket{n},\forall n\ge4$, can be coupled to distinct frequency bands. In this context, the RARE is capable of concurrently detecting communication signals at low ($<6{\rm GHz}$) or  medium bands ($7{\rm GHz}-24{\rm GHz}$) to ensure \emph{coverage}. Simultaneously, it can also capture sensing signals operating in the mmWave and THz bands to perform sensing tasks with varying \emph{granularity} requirements, ranging from ToA estimation to heartbeat detection. It is amazing that all these S\&C services are achieved by a single RARE device, which is nearly impossible for classic receivers due to the largely separated frequencies. 

Fig.~\ref{img:MSAC_performance} compares the MSAC performance of the RARE and classic receivers. We consider a dual-band S\&C system. A sub-6G device transmits communication signals operating at $3.213\:{\rm GHz}$ with an instantaneous bandwidth of $500\:{\rm kHz}$. Simultaneously, a mmWave device transmits sensing signals operating at $30.618\:{\rm GHz}$ with a bandwidth of $100\:{\rm kHz}$. Similar to \cite{QuanSense_Zhang2023}, the sensing signal is reflected by a target for detecting its vibration. 
A single RARE is deployed to detect these dual-band signals. The Rydberg states are configured as $\ket{3} = 60D_{5/2}$, $\ket{4} = 61P_{3/2}$, and $\ket{5} = 62P_{3/2}$, respectively. The electron transition $60D_{5/2}\rightarrow 61P_{3/2}$ has a dipole moment of $2.04\times10^{-26}\:{\rm C\cdot m}$ and resonates with the $3.213\:{\rm GHz}$ communication signal, while the transition $60D_{5/2}\rightarrow 62P_{3/2}$ has a dipole moment of $6.24\times10^{-27}\:{\rm C\cdot m}$ and resonates with the $30.618\:{\rm GHz}$ sensing part. 
For comparison, two classic receivers are also deployed to independently complete the aforementioned S\&C tasks, denoted as ``CR1" for the communication service and ``CR2" for the sensing. The transmission power increases from $-10\:{\rm dBm}$ to $10\:{\rm dBm}$. As observed
in Fig.~\ref{img:MSAC_performance}, the spectrum efficiency in the communication service achieved by 
RARE is $\sim2.4{\rm bps/Hz}$ higher than that of CR1 under Rayleigh fading channels.
Additionally, the normalized mean square error (NMSE) of vibration sensing achieved by the RARE improves
that achieved by CR2 by $\sim7.2\:$dB. Notably, these performance gains are accomplished by a single RARE without requiring additional RF circuits.
 
\begin{figure}
   \centering
   \includegraphics[width=3.5in]{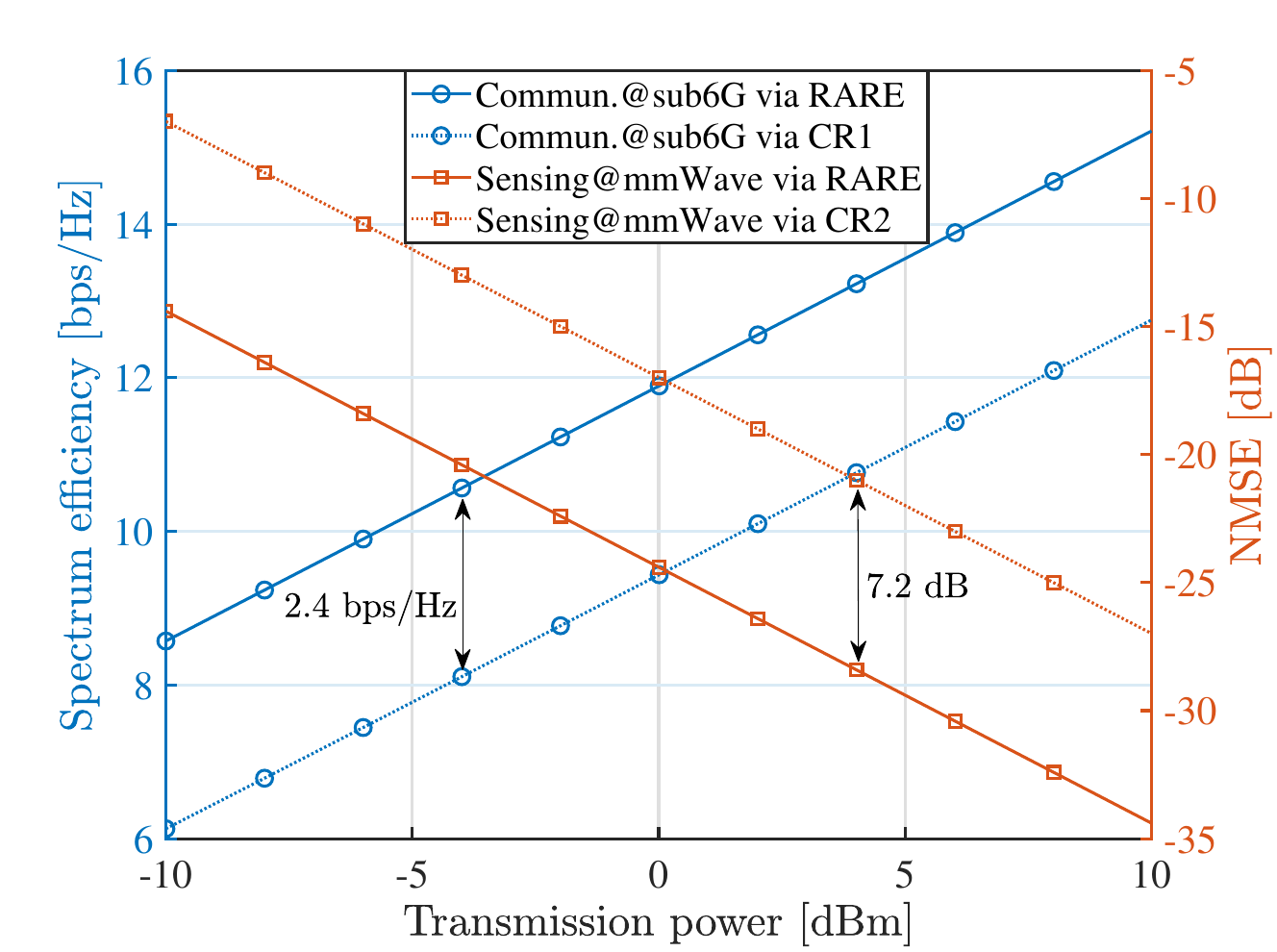}
   \vspace*{-1em}
   \caption{Comparison of MSAC performance between a single RARE and two classic receivers.   
   }
   \label{img:MSAC_performance}
   \vspace*{-1em}
\end{figure}

}

\section{Future Research Directions}
Despite the rapid development of RARE, its integration with practical communication systems remains at a nascent stage. Numerous physical-layer research problems deserve in-depth investigation to unleash the full potential of RAREs. 

\begin{itemize}
    \item {\bf More accurate transmission model.} The current signal transmission model of RARE primarily considers the interaction between Rydberg atoms and EM waves. In practice, Rydberg-Rydberg interactions can significantly influence the dynamics of quantum states. A comprehensive study of these interactions is essential to enhance the accuracy of the RARE signal transmission model.
    \item {\bf Joint transmitter and RARE optimization.} As RARE modifies the signal transmission model from linear to nonlinear, traditional signal processing methods become suboptimal. It is essential to develop new transmission strategies tailored for both the RARE and transmitter, including modulation, precoding, channel estimation, synchronization, interference mitigation, and so on. 
    \item {\bf  Holographic atomic-MIMO communications.} Conventionally, the performance of holographic MIMO---where massive antennas are deployed within a compact area---is limited by the mutual coupling between antenna elements. Fortunately, the holographic atomic-MIMO technique can overcome this limitation, as Rydberg atoms in different laser streams minimally influence each other.
    \item {\bf  Data-driven RARE.} To date, the quantum dynamics of Rydberg atoms under complicated EM environments, such as twisted radiowaves and broadband signals, remains largely unexplored due to the intricate Lindblad master equations. Data-drive methods, including physics-informed neural networks and foundation models, offer promising solutions for understanding and optimizing these RARE systems~\cite{liu_deep_2022}. 
    \item  {\bf Physical layer security.} RARE holds significant potential for enhancing physical layer security. Leveraging  abundant energy levels, RARE systems can flexibly switch their listening frequency bands. This capability enables high-speed and  ultra-broadband frequency-hopping communication, effectively protecting wireless data privacy.
    \item {\bf Satellite communication and low-altitude economy.} Another promising application of RARE is in supporting satellite communication and the low-altitude economy. The high sensitivity of RARE not only enables the expansion of satellite signal coverage but also demonstrates the potential to enhance the precision of sensing and navigation for unmanned aerial vehicles.
\end{itemize}

\section{Conclusions}
%

This article has provided an overview of RARE-aided wireless communications. Representing a paradigm shift in receiver architecture, RAREs detect EM signals via quantum phenomena, contributing to improved sensitivity and broadened detectable frequency range. RAREs can seamlessly integrate with both classic techniques, such as FDM and MIMO, as well as quantum techniques, like the many-body effect, to enhance communication performance. 
The potential of RARE in MSAC systems was revealed through experimental results. 
Finally, we have highlighted several promising future research directions on RARE.


	\section*{Acknowledgment}
    This work was supported in part by the National Natural Science Foundation of China (No. 624B2123), in part by the Research Grants Council of the Hong Kong Special Administrative Region, China under a fellowship award (HKU RFS2122-7S04), NSFC/RGC CRS (CRS\_HKU702/24), the Areas of Excellence scheme grant (AoE/E-601/22-R), Collaborative Research Fund (C1009-22G), and the Grant 17212423, and in part by the Shenzhen-Hong Kong-Macau Technology Research Programme (Type C) (SGDX20230821091559018).

\bibliographystyle{IEEEtran}
\bibliography{Reference.bib}

	\section*{Biographies}{
		{\textbf{Mingyao Cui} is a PhD candidate at The University of Hong Kong, Hong Kong.}
		\par \quad \par
        {\textbf{Qunsong Zeng} is a research assistant professor with The University of Hong Kong, Hong Kong.}
        		\par \quad \par

        {\textbf{Kaibin Huang} is a professor with The University of Hong Kong, Hong Kong.}
		\par \quad \par

\end{document}